%% file: CrossASRv2.tex
\documentclass[sigconf]{acmart}
\acmConference[ESEC/FSE 2021]{The 29th ACM Joint European Software Engineering Conference and Symposium on the Foundations of Software Engineering}{23 - 27 August, 2021}{Athens, Greece}

\usepackage{comment}
\usepackage{balance}
\usepackage{listings}
\usepackage{multirow}
\usepackage[ruled,linesnumbered]{algorithm2e}
\usepackage{bm}
\usepackage{url}
\SetAlFnt{\small}
\SetAlCapFnt{\small}
\SetAlCapNameFnt{\small}
\usepackage{algorithmic}
\algsetup{linenosize=\small}
\SetKwInput{KwInput}{Input}                % Set the Input
\SetKwInput{KwOutput}{Output}              % set the Output

\newboolean{showcomments}
\setboolean{showcomments}{true}
\ifthenelse{\boolean{showcomments}}
{ }
\ifthenelse{\boolean{showcomments}}
{ }
\ifthenelse{\boolean{showcomments}}
{ }
\AtBeginDocument{%
  \providecommand\BibTeX{{%
    \normalfont B\kern-0.5em{\scshape i\kern-0.25em b}\kern-0.8em\TeX}}}

\usepackage{tcolorbox}

\setcopyright{acmcopyright}
\copyrightyear{2021}
\acmYear{2021}
\acmDOI{10.1145/1122445.1122456}

\acmBooktitle{Woodstock '18: ACM Symposium on Neural Gaze Detection,
  June 03--05, 2018, Woodstock, NY}
\acmPrice{15.00}
\acmISBN{978-1-4503-XXXX-X/18/06}

\begin{document}
\def \toolname{{\em CrossASR++} }
\def\code#1{\texttt{\small{#1}}}

\title{\toolname{}: A Modular Differential Testing Framework for Automatic Speech Recognition}

\author{Muhammad Hilmi Asyrofi, Zhou Yang and David Lo}

\affiliation{%
  \institution{School of Computing and Information Systems}
  \country{Singapore Management University}
}
\email{{mhilmia, zyang, davidlo}@smu.edu.sg}

\renewcommand{\shortauthors}{Asyrofi and Yang, et al.}

\begin{abstract}
  Developers need to perform adequate testing to ensure the quality of Automatic Speech Recognition (ASR) systems. However, manually collecting required test cases is tedious and time-consuming. Our recent work proposes CrossASR, a differential testing method for ASR systems. This method first utilizes Text-to-Speech (TTS) to generate audios from texts automatically and then feed these audios into different ASR systems for cross-referencing to uncover failed test cases. It also leverages a failure estimator to find failing test cases more efficiently. Such a method is inherently self-improvable: the performance can increase by leveraging more advanced TTS and ASR systems. So in this accompanying tool demo paper, we devote more engineering and propose {\em CrossASR++}, an easy-to-use ASR testing tool that can be conveniently extended to incorporate different TTS and ASR systems, and failure estimators. We also make \toolname chunk texts from a given corpus dynamically and enable the estimator to work in a more effective and flexible way. We demonstrate that the new features can help \toolname discover more failed test cases. Using the same TTS and ASR systems, \toolname can uncover $26.2\%$ more failed test cases for 4 ASRs than the original tool. Moreover, by simply adding one more ASR for cross-referencing, we can increase the number of failed test cases uncovered for each of the 4 ASR systems by $25.07\%$, $39.63\%$, $20.95\%$ and $8.17\%$ respectively. We also extend \toolname with 5 additional failure estimators. Compared to worst estimator, the best one can discover $10.41\%$ more failed test cases within the same amount of time. 
  \newline Demo: \url{https://youtu.be/ddRk-f0QV-g} 
  \newline Tool: \url{https://github.com/soarsmu/CrossASRplus}  

\end{abstract}





\maketitle

\input{sections/intro.tex}
\input{sections/tool.tex}
\input{sections/usage.tex}

\input{sections/experiment.tex}
\input{sections/related_work.tex}
\input{sections/conclusion.tex}


% \begin{acks}
%   We thank the anonymous reviewers for their insightful comments. This research is supported by the Ministry of Education, Singapore, under its Academic Research Fund Tier 2 (Award No.: MOE2019-T2-1-193).
% \end{acks}

\balance
\bibliographystyle{ACM-Reference-Format}
\bibliography{reference}

\appendix

\end{document}

%% file: sections/intro.tex
\section{Introduction}
Automatic Speech Recognition (ASR), an essential technique supporting human-computer interaction, has been widely applied in modern life. Due to its ubiquity and importance, ensuring the quality of ASR systems is important. In software engineering, testing is a common practice to reveal defects, which can then be fixed to improve the quality of software products. Intuitively, a test case for an ASR system is straightforward: one piece of audio (input) and the corresponding transcription (oracle). However, manually curating these test cases requires significant human effort and time~\cite{115546}. As a result, researchers have developed automated test generation techniques to reduce the expense for testing ASR systems and helping uncover failures at an early stage. 

Such works can be divided into two branches: metamorphic testing and differential testing. The former is built on a basic assumption that adding a subtle disturbance to a piece of audio should not change the recognized transcript produced by an ASR system \cite{Du2018DeepCruiserAG, 8424625, Schnherr2019AdversarialAA, Qin2019ImperceptibleRA,Khare2018AdversarialBA,Abdullah2019HearE,8844615}. Work from the other branch, CrossASR \cite{Asyrofi2020CrossASR} that we recently introduce, employs Text-to-Speech (TTS) engines to synthesize test inputs from a text corpus. Then, it performs black-box differential testing on ASR systems by cross-referencing multiple ASRs to detect different transcriptions produced by the ASRs. If the transcribed text recognized by the ASR system under test does not match the text input to TTS while another ASR’s transcribed text does, this input audio is viewed as a failed test case. To increase its efficiency, CrossASR also employs a failure estimator to select texts from the input corpus that are more likely to become failed test cases.

Although these research works and prototypes demonstrate great potential to test ASR systems automatically, the community still lacks an easy-to-use tool that leverages state-of-the-art techniques. Besides, new TTS and ASR systems are continuously emerging and evolving. For example, wav2letter++ \cite{8683535}, an ASR system used in the CrossASR experiments, is no longer maintained (migrated to another platform) \cite{wav2letter_repo} and additional ASR systems, e.g. Wav2Vec2 \cite{baevski2020wav2vec}, are newly proposed. CrossASR can inherently improve its ability to uncover failed test cases for a system under test by utilizing more TTS and ASR systems for cross-referencing. These facts motivate us to devote more engineering effort and propose \toolname in this paper. We modularize the original tool so it can be easily extended to support additional TTS and ASR systems to achieve better performance.
Besides, the original CrossASR adopts a {\em static chunking} strategy: splitting an input corpus into a fixed number of batches. It analyzes only one batch in each iteration (with a specific timeout period). In each iteration, the visibility of the failure estimator is static and limited: it can only access texts in one batch to prioritize them (based on their estimated likelihood to uncover failures). To address this limitation, we design \toolname with dynamic chunking and flexible visibility. In each iteration, the failure estimator can produce estimates for more texts, and \toolname can prioritize which texts to be run first (to uncover failures) among a larger and dynamic pool of texts (that changes with each iteration). Our evaluation results show that CrossASR++ outperforms CrossASR by uncovering many more failures given a fixed time budget.

The rest of this paper is organized as follows. Section \ref{tool_design} describes the workflow, main features and a usage example. In Section \ref{experiment}, we evaluate the effectiveness of \toolname and its features in uncovering failures. Section \ref{rel_work} discusses some related work. Finally, we conclude the paper and present future work in Section \ref{conclusion}.

%% file: sections/tool.tex
\section{\toolname}
\label{tool_design}
\subsection{Workflow}
\label{workflow}
This section discusses the workflow of \toolname and Algorithm \ref{algo1:crossasr-workflow} illustrates this workflow.
Given a text corpus, \toolname processes the corpus gradually in multiple iterations. We first initialize some important variables (Line 1 - 2).
At the start of the first iteration, no test cases are processed yet. So the algorithm will skip the body of the {\verb|if|} statement (Line 5-8) and keep the text batch selected in Line 2. Then, TTS is used to generate an audio file for each piece of selected text (Line 9). In the next step, we use all ASR systems (including the System Under Test (SUT)) to convert the audio files back to texts (Line 11), and then we perform cross-reference (Line 12). For a piece of audio and its corresponding transcription, there are three possible situations: 1) if the SUT can recognize it successfully, i.e. the transcript matches the original text, we take it as a {\em successful test case}. 2) If the SUT fails to recognize the audio file successfully, but at least one of the other ASR systems succeeds, we take this audio file (and the corresponding input text) as a {\em failed test case} for the SUT. 3) If all the ASR systems cannot recognize the audio correctly, we call it an {\em indeterminable test case} because a TTS may generate an invalid audio. After cross-referencing, we store uncovered failed test cases (Line 13) and other test cases (Line 14).

To enable \toolname to identify more failed test cases given a time budget, we make use of a failure estimator. At the start of each iteration except the first one, we use failed test cases, successful test cases, and indeterminable test cases so far to train a failure estimator (Line 6). This estimator can estimate the probability of a piece of text leading to a failed test cases. We rank the texts to be chosen in the iteration and prioritize texts with a higher ranking to be processed (until the time budget of the iteration has been expended) (Line 7). 

By default, the time budget for each iteration is one hour. If the time budget has been expended in an iteration, \toolname will proceed to the next iteration. So it is possible that \toolname only processes a part of texts in a batch.

\begin{algorithm}[t]
  \caption{\toolname Workflow
  } 
  \label{algo1:crossasr-workflow}
  \SetAlgoLined
  \KwInput{{\em corpus}: a list of texts}
  \KwOutput{{\em failed tests} : generated failed test cases}
  {\em failed\_tests, other\_tests} = None, None\;
  {\em texts} = {\em getFirstBatch(corpus)}\;
  \While{not exceed maximal iteration}{
    {\#\# new text selection}\;
    \If{{\em $\mathit{failed\_tests}$} not None}{
      {\em estimator.train(failed\_tests, other\_tests)} \;
      {\em texts} = {\em estimator.select(getNextBatch(corpus))}\;
    }

    {\em audios} = {\em tts.generateAudio(texts)}\;
    {\#\# cross-referencing process}\;
    {\em transcriptions} = {\em asrs.recognizeAudio(audios)}\;
    {\em f\_tests, o\_tests} = {\em crossReference(transcriptions, texts)}\;
    {\em failed\_tests.append(f\_tests)}\;
    {\em other\_tests.append(o\_tests)}\;

  }
  \algorithmicreturn{ $\mathit{failed\_tests}$}
\end{algorithm}

\subsection{Extensibility} \label{extensibility}
We devote more engineering effort to enhancing the extensibility of the original CrossASR. We believe such extensibility can make \toolname self-improvable: the performance can increase by leveraging more advanced TTS and ASR systems. Theoretically, if no timeout is set for each iteration, adding \toolname with more ASR systems for cross-referencing never decreases the number of failed test cases uncovered. The intuition is that newly added ASR systems may turn indeterminable test cases into failed test cases while the original successful and failed test cases remain. As reported in our prior paper \cite{Asyrofi2020CrossASR}, TTS systems also differ in their ability to help in finding failed test cases. The main reason is that more advanced TTSes generate fewer invalid audios that are more likely to be indeterminable test cases. As a result, using better TTSes can improve \toolname. The failure estimator can estimate the probability that a text leads to a failed test case. Intuitively, a better failure estimator can help \toolname find more texts that result in failed test cases. The analysis above motivates us to enhance the extensibility of the original CrossASR by introducing a modular design to CrossASR++. 

We discard the implementation of the original CrossASR as it serves more as a prototype to demonstrate the viability of the research idea rather than a tool that others can easily use and expand. So we implement all necessary processes presented in Algorithm \ref{algo1:crossasr-workflow} and pay attention to its extensibility. Extensibility is mainly enabled by modeling a TTS, ASR, and failure estimator with interfaces, i.e. abstract base classes. Users can add a new TTS, a new ASR or a new failure estimator by simply inheriting the base class and implementing necessary methods.

We have 3 base classes, i.e. {\verb|ASR|}, {\verb|TTS|}, and {\verb|Estimator|}. When inheriting from each class, users need to specify a name in the constructor. This name will be associated with a folder for saving the audio files and transcriptions. Thus, we require that each derived class has a unique name. When inheriting {\verb|ASR|} base class, users must override the {\verb|recognizeAudio()|} method which takes an audio as input and returns recognized transcription. % The added ASR systems can be used as the system under test or for cross-referencing purpose.
TTS and failure estimator can be added similarly. In {\verb|TTS|} base class, the method {\verb|generateAudio()|} must be overrided by derived classes. This method converts a piece of text into audio. In {\verb|Estimator|} base class, methods {\verb|fit()|} and {\verb|predict()|} must be overrided by derived classes. These methods are used for training and predicting, respectively. 

As default setting of \toolname, we have incorporated some latest components.  The suppported TTSes are Google Translate's TTS~\cite{google_tts}, ResponsiveVoice~\cite{rv_tts}, Festival~\cite{festival_tts}, and Espeak~\cite{espeak_tts}. The supported ASRs are DeepSpeech~\cite{Hannun2014DeepSS}, DeepSpeech2~\cite{pmlr-v48-amodei16}, Wit~\cite{wit_asr}, and wav2letter++~\cite{8683535}. \toolname supports any transformed-based classifier available at HuggingFace~\cite{huggingface_repo}. \toolname can be easily extended to leverage more advanced tools in the future.

\subsection{Dynamic Chunking and Visibility}
The original CrossASR adopts a {\em static chunking} strategy. Before processing the corpus, it splits the corpus into a fixed number of batches and assigns each batch to one iteration. When processing texts in each iteration, CrossASR can only access a limited number of texts assigned in the text batch. If it reaches the time limit, it will discard all the unprocessed texts in the text batch. Also, the estimator can only predict the failure probability of texts in that batch as well. There are two main drawbacks of this {\em static chunking} strategy. First, the unprocessed texts can also be failed test cases. Thus discarding them will decrease the number of failed test cases. Second, the visibility of the estimator is limited. The limited visibility means that the estimator can only access a small number of texts to prioritize. No matter how effective the estimator is, it can only give a higher rank to a few texts that are likely to be failed test cases. 

We design \toolname with {\em dynamic chunking} and more flexible visibility. We allow users to adjust the visibility of the failure estimator, and the text batches are no longer split statically before processing. After ranking texts according to their failure probabilities, \toolname processes the texts starting from the one with the highest failure probability estimate. There is still a timeout (or time budget) for each iteration, but when the timeout is reached, \toolname appends unprocessed texts to the next iteration rather than simply discards them. In other words, \toolname chunks texts in each iteration dynamically.

%% file: sections/usage.tex
\subsection{Tool Usage}
\subsubsection{Installation}

\toolname can be installed via a simple command: {\verb|pip install crossasr|}. The components that \toolname relies on, e.g. TTS, ASR systems and estimators, can also be easily install and incorporated into \toolname as described in Section \ref{extensibility}. Please refer to the Github repository for more information.

\subsubsection{Configuration and Execution}

After adding ASRs, TTSes and estimators, we can run \toolname with some configuration parameters. Users need to set {\verb|tts|}, {\verb|asrs|}, {\verb|target_asr|} and {\verb|estimator|} parameters to let \toolname know the TTS to generate audios, ASRs for cross-reference, the ASR under test and the failure estimator to use. Each of these parameters takes a string or a list of strings. % \mh{david ask to delete this} These strings must be selected from the names specified in Section \ref{extensibility}.

Other important parameters are {\verb|num_iteration|}, {\verb|time_budget|}, {\verb|text_batch_size|}, and {\verb|recompute|}. \toolname runs in multiple iterations as specified by {\verb|num_iteration|}. The timeout for each iteration is limited to {\verb|time_budget|} measured in second. The number of texts can be processed in each iteration, i.e. the visibility, can be set by specifying {\verb|text_batch_size|}. Users can adjust these parameters according to their situation. For example, users with a sufficient time can increase the {\verb|time_budget|} and {\verb|num_iteration|} to generate more failed test cases. In the Github repository, we also provide all the TTS-generated audios we use. Users can set {\verb|recompute|} as {\verb|false|} to use them directly. % \mh{david ask to delete} without the need to add TTSes that are normally commercial and may have API call limitations. 

All of the configuration parameters are saved at {\verb|config.json|}. We provide an example script {\verb|test_asr.py|}. Users can execute ASR testing with command {\verb|python test_asr.py config.json|}.

% After adding neccessary ASRs, TTSes and estimators into the tool, we can use {\verb|config.json|} to specify the parameters needed to run {\em CrossASR++}. These parameters are presented in Table \ref{tab:config-parameter}. Here we only discuss some of them.

% First, we need to modify {\verb|tts|}, {\verb|asrs|}, {\verb|target_asr|} and {\verb|estimator|} to let \toolname know the TTS to generate audios, ASRs for cross-reference, the ASR system under test and the failure estimator to use. The four parameters take a string or a list of strings. These string must be selected from the names specified in Section \ref{extensibility}.

% \begin{table}
%   \caption{Configuration Parameters.}
%   \label{tab:config-parameter}
%   \vspace{-0.3cm}
%   \begin{tabular}{rl}
%     \toprule
%     \textbf{Parameter} & \textbf{Description}\\
%     \midrule
%     \code{seed} & random seed for reproducibility\\
%     \code{output\_dir} & directory to save the experiment output \\ \code{tts} & TTS used \\
%     \code{asrs} & list of ASRs for cross-reference \\
%     \code{target\_asr} & ASR under test \\
%     \code{estimator} & estimator used to speed up the process \\
%     \code{recompute} & recompute the experiment from scracth \\
%     \code{num\_iteration} &  number of iterations \\
%     \code{time\_budget} & allocated time for each iteration\\
%     \code{text\_batch\_size} & number of texts processed in each iteration \\
%     \bottomrule
%   \end{tabular}
%   \vspace{-0.3cm}
% \end{table}

%% file: sections/experiment.tex
\section{Experiments}
\label{experiment}
\begin{table}[!t]
  \centering
  \caption{Results for \toolname and the original CrossASR.}
  \vspace{-0.2cm}
  \begin{tabular}{lccccc}
    \toprule
    \multirow{2}{*}{\textbf{Config}} &
    \multicolumn{4}{c}{\textbf{\# Failed Test Case}}  & 
    \multirow{2}{*}{\textbf{Total}} \\
      & \textbf{DS} & \textbf{DS2} & \textbf{W2L} & \textbf{Wit} \\
    \midrule
    CrossASR & 266 & 171 & 391 & 704 & 1,532 \\
    CrossASR++ & 319 & 217 & 420 & 978 & 1,934 \\
    \bottomrule
  \end{tabular}
  \vspace{-0.3cm}
  \label{tab:exp1}
\end{table}
In this section, we evaluate \toolname using the Europarl dataset -- the same corpus was used to evaluate CrossASR as reported in \cite{Asyrofi2020CrossASR}. After removing duplicates and dropping empty texts, we randomly pick 20,000 texts for our evaluation. To measure how \toolname performs in finding failed test cases, we answer the following two research questions. For both RQs, we use ResponsiveVoice as TTS, set timeout as 1 hour and run for 5 iterations (with a time budget of 1 hour for each iteration).

\vspace{0.2cm} \noindent \textbf{RQ1.} \textit{How many failed test cases can CrossASR++ find?}

To answer RQ1, we run \toolname with four ASR systems under two configurations. The first configuration uses static chunking and the {\verb|albert-base-v2|} estimator (the setting used by CrossASR), while the second configuration uses dynamic chunking, sets visibility to 1,200, and uses a more advanced estimator ({\verb|facebook-bart-base|}). These two configurations are helpful for comparing the superiority of \toolname over CrossASR. Table~\ref{tab:exp1} illustrates the results of running the tool using the two configurations. Each column represents the number of failed test cases uncovered for the system under test (SUT) when we use one ASR as the SUT and the remaining 3 ASRs for cross-referencing. The result shows that for each SUT, \toolname can help find more failed test cases. In total, \toolname can find 26.2\% more failed test cases than the original tool, which demonstrates significant improvements to the original tool.

\vspace{0.2cm} \noindent \textbf{RQ2.} \textit{To what extent do enhanced features help find more failed test cases?}

We enhance \toolname with three main features: (1) extensibility to use more ASR, (2) dynamic chunking and flexible visibility of estimator, and (3) extensibility to use advanced estimators. In this RQ, we select the second configuration from RQ1 as the base configuration. By means of an ablation study, we explore how these three features contribute to the performance of CrossASR++; specifically, we modify only one parameter related to one of the features each time and keeping the other parameters untouched.

First, we analyze the effect of extensibility to use more ASRs. Table~\ref{tab:failed-test-cases} illustrates the changes in number of failed test cases when adding one more ASR ({\verb|Wav2Wec2|}) into the system. We can observe that by simply adding one more ASR for cross-referencing can increase the number of failed test cases uncovered for 4 SUTs by $25.07\%$, $39.63\%$, $20.95\%$ and $8.17\%$ respectively. The observation aligns with our analysis in Section \ref{extensibility}.

\begin{figure}[t]
    \centering
    \vspace{0.2cm}
    \includegraphics[width=0.8\linewidth]{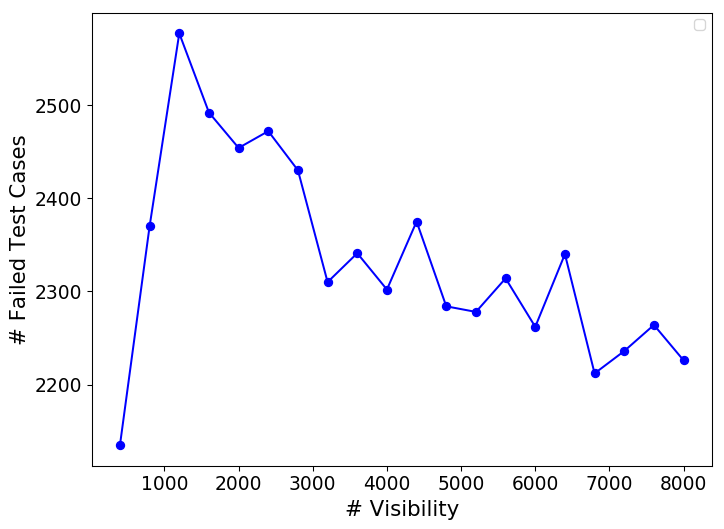}
    \vspace{-0.3cm}
    \caption{Results when using different visibility settings.}
    \label{fig:rq3-trend}
\end{figure}

Then, we analyze how dynamic chunking strategy and visibility affect the effectiveness of {\em CrossASR++}. We use the first configuration from RQ1 and only change static chunking to dynamic chunking. We found that the total number of failed test cases increases from 1,532 to 1,567, emphasizing that dynamic chunking is better. Using the second configuration from RQ1, we also examine how \toolname performs under different settings of visibility. Figure \ref{fig:rq3-trend} illustrates the trend of number of failed test cases uncovered when visibility is varied from 400 to 8,000. It can be observed that early in this range (400 to 8,000), using larger visibility can achieve better results than using the original setting, 400. Moreover, there is a peak for visibility's impact, and increasing visibility further reduces the number of failed test cases generated. According to our experiment, 1,200 is an ideal setting for visibility, which can increase the total number of failed test cases by $16.34\%$ compared to setting it to 400.

\begin{table}[!t]
  \centering
  \caption{Adding one more ASR to the system.}
  \vspace{-0.2cm}
  \begin{tabular}{cccccccc}
    \toprule
    \multirow{2}{*}{\textbf{ASR}} & \multirow{2}{*}{\textbf{Visibility}} &
    \multicolumn{5}{c}{\textbf{\# Failed Test Case}}  &
    \multirow{2}{*}{\textbf{Total}} \\
      & & \textbf{DS} & \textbf{DS2} & \textbf{W2L} & \textbf{Wit} & \textbf{W2V} \\
      \midrule
      4 & 1,200 & 319 & 217 & 420 & 978 & - & 1,934 \\
      5 & 1,200 & 399 & 303 & 508 & 1,058 & 309 & 2,577 \\
      % \midrule
      % 4 ASRs & 450 & 315 & 506 & - & 305 & 1,576 \\
      % 4 ASRs & 480 & 336 & - & 1,674 & 328 & 2,818 \\
      % 4 ASRs & 265 & - & 350 & 1,106 & 165 & 1,886 \\
      % 4 ASRs & - & 251 & 429 & 1,263 & 245 & 2,188 \\
  \bottomrule
  \end{tabular}
  \vspace{-0.3cm}
  \label{tab:failed-test-cases}
\end{table}

% \begin{table}[!t]
%     \centering
%     \caption{The number of failed test cases generated after adding an ASR.}
%     \vspace{-0.2cm}
%     \begin{tabular}{cccccccc}
%       \toprule
%       \multirow{2}{*}{\textbf{ASR}} & \multirow{2}{*}{\textbf{Visibility}} &
%       \multicolumn{5}{c}{\textbf{\# Failed Test Case}}  &
%       \multirow{2}{*}{\textbf{Total}} \\
%         & & \textbf{DS} & \textbf{DS2} & \textbf{W2L} & \textbf{Wit} & \textbf{W2V} \\
%         \midrule
%         4 & 1,200 & 319 & 217 & 420 & 978 & - & 1,934 \\
%         5 & 1,200 & 399 & 303 & 508 & 1,058 & 309 & 2,577 \\
%         5 & 400 & 315 & 210 & 367 & 1,096 & 227 & 2,215 \\

%         % \midrule
%         % 4 ASRs & 450 & 315 & 506 & - & 305 & 1,576 \\
%         % 4 ASRs & 480 & 336 & - & 1,674 & 328 & 2,818 \\
%         % 4 ASRs & 265 & - & 350 & 1,106 & 165 & 1,886 \\
%         % 4 ASRs & - & 251 & 429 & 1,263 & 245 & 2,188 \\
%     \bottomrule
%     \end{tabular}
%     \vspace{-0.3cm}
%     \label{tab:failed-test-cases}
% \end{table}

Lastly, we compare the impacts of using different estimators. We fix other parameters and only replace failure estimators used. We try no estimator and 6 different transformer-based estimators. Table~\ref{tab:performance-improvement} shows the total number of failed test cases when using different estimators. We find that the best result comes from a recently-released model \verb|facebook-bart-base|, which is the default setting of {\em CrossASR++}. We can observe that estimators can make a difference: the best estimator helps find $10.41\%$ more failed test cases than the worst estimator. 

The above analysis shows that all the three enhanced features we incorporate into \toolname can help boost its performance in finding more failed test cases.

\begin{table}[!t]
  \caption{Results when using different estimators.}
  \label{tab:performance-improvement}
  \begin{tabular}{lc}
    \toprule
    \textbf{Estimator} & \textbf{\# Failed Test Cases} \\
    \midrule
    No estimator & 1,574 \\
    bert-base-uncased & 2,334 \\
    albert-base-v2 & 2,419 \\
    distilbert-base-uncased & 2,446 \\
    xlnet-base-cased & 2,495 \\
    roberta-base & 2,559 \\
    \textbf{facebook-bart-base} & \textbf{2,577} \\
  \bottomrule
\end{tabular}
\end{table}

%% file: sections/related_work.tex
\section{Related Work}
\label{rel_work}
This section briefly introduces works related to testing ASR systems. We divide current methods to test ASR into two branches: metamorphic testing and differential testing. Most of works fall into the first branch. Researchers from SE communities \cite{Du2018DeepCruiserAG, guo2021rnntest}, follow the conventional software testing principles: define some coverage criteria in ASR systems and generate metamorphic transformations guided by these coverage criteria. Then, they apply transformations to original audios to derive mutants that are expected to have the same recognized transcripts as the original ones. In the AI community, researchers often generate such transformations adversarially by using optimization-based methods \cite{8424625, Schnherr2019AdversarialAA, Qin2019ImperceptibleRA,Khare2018AdversarialBA,Abdullah2019HearE,8844615}.

The other branch, differential testing \cite{df_testing}, provides the same input to different softwares with the same function and to see whether they produce different outputs. CrossASR \cite{Asyrofi2020CrossASR}, which is our prior work, is the first to apply differential testing to ASR systems. CrossASR employs Text-to-Speech (TTS) engines to synthesize test inputs from texts. Then it performs black-box differential testing on ASR systems by cross-referencing multiple ASRs to detect different transcriptions recognized by ASRs. If the transcribed text recognized by the ASR system under test does not match the text input to a TTS, while another ASR’s transcribed text does, this input audio is viewed as a failed test case. To increase its effectiveness to find failed test cases, CrossASR utilizes a failure estimator that selects texts that are more likely to lead to failed test cases.

%% file: sections/conclusion.tex
\section{Conclusion and Future Work}
\label{conclusion}

This paper presents \toolname, an extensible tool that performs black-box differential testing on ASR systems. \toolname can be conveniently extended to incorporate different TTS and ASR systems, and failure estimators. The extensibility allows \toolname to flexibly utilizes more tools to boost its performance on uncovering failed test cases. In the default package of {\em CrossASR++}, we incorporate 4 TTS and 5 ASR systems, as well as 6 failure estimators. Our evaluation results show that \toolname outperform CrossASR by revealing 26.2\% more failed test cases. In addition, we find that the new features introduced by \toolname help discover more failed test cases. For example, by simply adding one more ASR system for cross-referencing, the number of failed test cases for each of the 4 ASRs under test increases by $25.07\%$, $39.63\%$, $20.95\%$ and $8.17\%$ respectively. The number of failed test cases also increases by increasing the visibility until some point. We also find that leveraging a more advanced failure estimator can help \toolname achieve better performance. We encourage practitioners and researchers to augment \toolname with more TTS and ASR systems and failure estimators.